\documentclass[12pt]{article}
\usepackage{amsthm}
\usepackage{epsfig}   
\usepackage{amssymb}
\input{epsf}

\def\be{\begin{equation}}
\def\bea{\begin{eqnarray}}
\def\ee{\end{equation}}
\def\eea{\end{eqnarray}}

\def\apx{\approx}

\def\wtd{\widetilde}

\def\pt{\partial}

\def\ovl{\overline}

\def\pr{\prime}

\def\Gm{\Gamma}
\def\dt{\delta}
\def\Dt{\Delta}
\def\eps{\varepsilon}

\def\Ffi{\Phi}
\def\kp{\kappa}

\def\la{\lambda}
\def\Sg{\Sigma}
\def\sg{\sigma}

\def\th{\theta}

\def\om{\omega}

\def\sign{\mbox{sign}}

\def\const{\mbox{const}}

\def\vol{\mbox{vol }}

\def\ln{\mbox{ln}}
\def\vol{\mbox{vol }}

\def\dint{\displaystyle \int }

\def\dfrac#1#2{{\displaystyle {#1 \over #2}}}

\begin{document}

\begin{center}

{\LARGE \bf On passage through resonances in volume-preserving systems}\\

\vskip 10mm

{\Large D.L.Vainchtein$^{\left. 1,2 \right)}$, A.I.Neishtadt$^{\left. 2 \right)}$, and
I.Mezic$^{\left. 1 \right)}$}\\

\vskip 5mm

{\large $^{\left. 1 \right)}$ \it Department of Mechanical and Environmental Engineering
\\ University of California Santa Barbara, CA 93106, USA}\\

\medskip

{\large $^{\left. 2 \right)}$ \it Space Research Institute, ul.Profsouznaya, 84/32, Moscow, Russia,
GSP-7, 117997}\\

\end{center}

\date{}

\vskip 30mm

\begin{abstract}
Resonance processes are common phenomena in multiscale (slow-fast) systems. In the present paper we
consider capture into resonance and scattering on resonance in 3-D volume-preserving slow-fast
systems. We propose a general theory of those processes and apply it to a class of viscous
Taylor-Couette flows between two counter-rotating cylinders. We describe the phenomena during a
single passage through resonance and show that multiple passages lead to the chaotic advection and
mixing. We calculate the width of the mixing domain and estimate a characteristic time of mixing.
We show that the resulting mixing can be described using a diffusion equation with a diffusion
coefficient depending on the averaged effect of the passages through resonances.
\end{abstract}

\section{Introduction.}

Appearing in a variety of settings, multiscale systems present quite a challenge for direct
numerical simulations (see e.g. \cite{Jansson:2005}). Thus, a development of analytical tools is
rather important. For a wide class of low-dimensional multiscale systems (that are also called
slow-fast systems) where some variables change with a characteristic rate of order $1$, while the
other variables change with a rate of order of $\eps \ll 1$ the averaging method can be used for
approximate description of the dynamics (see e.g. \cite{AKN:1988}). A standard condition for
applicability of the averaging method is that the frequency of motion of the fast variables is
indeed of order $1$ everywhere in the space. The violation of the above condition near resonances
of the fast system may give rise to the following {\it single resonance phenomena}: first,
separatrix crossings or, second, scattering on resonance and capture into resonance. The
hydrodynamic flows involving the separatrix crossings were introduced in
\cite{BajerandMoffatt:1990,StoneNadimandStrogatz:1991} and later discussed from the theoretical,
numerical and experimental perspectives (see e.g.
\cite{WardandHomsy:2001,WardandHomsy:2003,Grigoriev:2005}). The theory of separatrix crossings for
such flows was developed in \cite{VVN:1996a,NeishtadtandVasiliev:1999}.

The flows where the chaotic advection is caused by scattering on resonance and capture into
resonance were introduced in \cite{CartwrightFeingoldandPiro:1995,CartwrightFeingoldandPiro:1996}.
The later works include \cite{Solomon:2003}. However, the corresponding theory, that gives a
quantitative description of either single crossings or the long-time dynamics was not developed.

There are two major objectives of the present paper. First, we propose a general theory of
scattering on resonance and capture into resonance in volume-preserving systems. In our study we
rely heavily on the close analogy between volume-preserving and Hamiltonian systems, for which the
theory of scattering on resonance and capture into resonance was developed in
\cite{Neishtadt:1999}. Second, we would like to demonstrate that the resonance phenomena can be
used in real settings to create flows with coexisting mixing and non-mixing (KAM) domains with the
respective size and mixing rates easily and independently manageable. In the process, we want to
emphasize that the presence of resonances is not a sufficient condition for chaotic advection and
to illustrate some of the necessary ingredients for the resonance phenomena to create the chaotic
advection.

The structure of the paper is as follows. In Sections~2 to 4 we discuss a general theory of
passages through resonances in near-integrable volume-preserving 3-D systems. In Sect.~2 we
introduce basic equations and study the unperturbed and the averaged systems. In Sect.~3 we
consider the structure of a resonance and discuss the breakdown of the method of averaging.
Section~4 contains a detailed description of the general properties of the streamlines' dynamics in
the vicinity of a resonance. Scattering on resonance and capture into resonance are discussed in
Subsects.~4.1 and 4.2, respectively. We show that both capture and scattering destroy the adiabatic
invariance and we obtain expressions for jumps of adiabatic invariants during a single passage. As
the magnitude of the jumps is very sensitive to initial conditions, in the case of multiple
passages through resonance these phenomena can be treated as random processes and we derive their
statistical properties. In Sect.~5 we apply the technique developed in the previous sections to a
particular problem of mixing between two counter-rotating cylinders. In Sect.~6 we discuss the long
time dynamics of streamlines and present results of numerical simulations. We show that multiple
passages through resonance lead to the destruction of adiabatic invariance and chaotic advection
and estimate characteristic time of mixing. And finally, we show that mixing can be described using
a single diffusion-type PDE, in which the diffusion coefficient is provided not by the molecular
diffusion, but by the averaged influence of the resonance processes. Thus, we call the resulting
diffusion the {\it adiabatic diffusion}.

\section{Main equations. The unperturbed and the averaged systems.}

In the first part of the paper (Sects.~2--4) we present a general theory of passages through
resonances in near-integrable volume-preserving 3-D systems. In other words, we study the flows
that are a superposition of an integrable base flow and a weak perturbation. While the inclusion of
the time-dependence adds to the variety of possible flow configurations (see e.g.
\cite{CartwrightFeingoldandPiro:1995,Solomon:2003}) and extends the range of admissible resonances,
we will see that our (time-independent) setting is sufficient to observe most of the related
phenomena and is better suited for analytical description.

In the present paper we consider the base flows belonging to a class of integrable 3-D flows that
possesses two invariants. Introducing these invariants as new variables we can write the evolution
equations in the following form:
\bea
\dot{I}_1 &=& 0,
\nonumber \\
\dot{I}_2 &=& 0,
\nonumber \\
\dot{\th} &=& \om\left( I_1, I_2 \right). \nonumber
\eea
In such flows almost all the streamlines are closed curves (except for those that connect the
stationary points). Each curve is defined by the values of $I_1$ and $I_2$ and we denote it as
$\Gm_{I_1, I_2}$. Such systems belong to the two actions and one angle class of dynamical systems,
where a single changing variable being a coordinate along a streamline.

A generic small perturbation causes the invariants of the base flow to change and a result flow has
the following form:
\bea
\dot{I}_1 &=& \eps v_1\left( I_1, I_2, \th\right),
\nonumber \\
\dot{I}_2 &=& \eps v_2\left( I_1, I_2, \th\right),
\label{vp1} \\
\dot{\th} &=& \om\left( I_1, I_2 \right) + \eps g\left( I_1, I_2 , \th\right). \nonumber
\eea
System (\ref{vp1}) is volume-preserving if the right hand side is divergence-free. For $0< \eps \ll
1$ flow (\ref{vp1}) possesses two time scales. The variable $\th$ is fast and the variables $I_1$
and $I_2$ are slow with a ratio of characteristic velocities being of order of $\eps$. System
(\ref{vp1}) is an example of so-called slow-fast systems. Therefore, in the first approximation we
can average velocity field (\ref{vp1}) over the fast variable $\th$. It is clear that this
approximation is valid everywhere except for a small part of the space where $\dot \th \apx 0$. The
averaging over $\th$ reduces (\ref{vp1}) to
\bea
\dot{I}_1 &=& \eps \frac1{2\pi}\dint_0^{2\pi} v_1\left( I_1, I_2, \th\right) \; d\th,
\nonumber \\
\label{vp2} \\
\dot{I}_2 &=& \eps \frac1{2\pi}\dint_0^{2\pi} v_2\left( I_1, I_2, \th\right) \; d\th. \nonumber
\eea
For the sake of brevity, in (\ref{vp2}) we used the same notation for $I_1$ and $I_2$ as in
(\ref{vp1}). System (\ref{vp2}) is 2-D volume-preserving and, therefore, Hamiltonian and
integrable. The Hamiltonian $\Ffi(I_1, I_2)$ is the flux of the perturbation vector ${\bf v} =
(v_1, v_2, g)$ across a surface $S$ spanning the streamline of the unperturbed system $\Gm_{I_1,
I_2}$. The quantity $\Ffi$ is an integral of the averaged system (\ref{vp2}) and it is an adiabatic
invariant of the exact system (\ref{vp1}) (see e.g. \cite{VVN:1996a,NeishtadtandVasiliev:1999}). If
$\dot{\th}$ does not vanish anywhere on a streamline, $\Ffi(I_1, I_2)$ would be conserved with the
accuracy of order $\eps$ over times of order $1/\eps$ (see c). Note, that the conservation of
$\Ffi$ is a direct consequence of the validity of averaging.

\section{Structure of the resonance and the breakdown of the method of averaging.}

It is widely known in the theory of dynamical systems that passages through the regions where the
frequencies of a given system satisfy the rational condition (in terms of (\ref{vp1}), $n_1 \eps
v_1 + n_2 \eps v_2 + n_3 (\om + \eps g) = 0$, where $n_i$'s are integer numbers) lead to chaotic
dynamics. As the ratio of characteristic frequencies in \ref{vp1}) is of order of $\eps$, for the
resonance to be of a low order we must have $\om = 0$. This is the only important resonance in the
systems similar to \ref{vp1}). In the process of motion some streamlines pass through the region
where $\dot \th \apx 0$, that we call a {\it resonance zone}. In the next two sections we develop a
general theory of resonance crossings in volume-preserving systems analogous to the one first
developed in \cite{Neishtadt:1997} and then adjusted in \cite{Neishtadt:1999} for Hamiltonian
systems. The equation $\om(I_1, I_2) = 0$ defines a 2-D surface in the original 3-D space (a curve
on the $(I_1, I_2)$ plane), that we call the {\it resonant surface}, or the {\it resonance}, and
denote by ${\mit R}$. In the vicinity of a resonance the variable $\th$ is not fast compared with
$I_1$ and $I_2$. Hence, we cannot expect averaged system (\ref{vp2}) to approximate the exact
system adequately. As a result, the value of the integral of the averaged system, $\Ffi$, may
change in the process of a passing through the vicinity of the resonance.

Qualitatively a single passage through the vicinity of a resonance can be described as follows. A
characteristic streamline of the exact system approaches the resonant zone with the value of $\Ffi$
oscillating with a small amplitude, $\sim \eps$, near some value $\Ffi^-$. When in the process of
the motion it arrives to the resonant zone, it is either captured into the resonance, or crosses
the resonant zone without being captured. (Actually, there is also some intermediate regime of
motion in the resonant zone, but it occurs for a very small measure of initial conditions; we will
not discuss it.) Phenomenologically, the difference between the two regimes of motion in the
resonant zone can be described as follows. In the case of capture, upon the arrival to the resonant
zone the phase $\th$ switches its behavior from rotation ($\th$ changes monotonically, a streamline
makes full circles around the axis) to oscillation ($\th$ changes between two values, $\dot{\th}$
changes the sign twice during each period). In the case of crossing the resonant zone without
capture, $\dot{\th}$ changes the sign only once in the resonant zone. As we show below, the two
phenomena have qualitatively different impact on the streamlines' behavior. After the passage
through the resonant zone (and far from the resonance) the value of $\Ffi$ oscillates near some
other value, $\Ffi^+$, again with a small amplitude $\sim \eps$.

In the rest of the present section we provide a qualitative description of these phenomena (without
any proofs) and return to a detailed quantitative description in the next section.

In the case of capture into resonance, upon the arrival to the resonant zone a phase point drifts
for a long time (of order $\sim 1/ \eps$) along the resonant surface. As a result, $\Ffi$ changes
by a value of order $1$. Among all the streamlines that arrive to the resonant zone during a given
time interval of order $\sim 1/ \eps$ only a small part, of order $\sim\sqrt{\eps}$, is captured.
Initial conditions for streamlines that are or are not captured are mixed. Therefore, it is
reasonable to consider capture as a probabilistic phenomenon.

The streamlines that cross the resonant zone without being captured typically pass through this
zone in time of order $\sim 1/\sqrt{\eps}$ (see \cite{Neishtadt:1997,Neishtadt:1999} for more
accurate estimates for Hamiltonian systems). The major resonant phenomenon for such streamlines is
the scattering on resonance. In this case the difference, $\Dt \Ffi = \Ffi^+ - \Ffi^-$, that is
typically of order $\sim\sqrt{\eps}$, is referred to as the amplitude of the scattering or, as
$\Ffi$ is the adiabatic invariant of the averaged system, as a {\it jump of the adiabatic
invariant}. This value is very sensitive to small changes of the initial conditions: a change of
the initial conditions far from a resonance by a quantity of order $\sim \eps$ leads to a big (of
order $\sim 1$ compared with the typical values) relative change in the amplitude of the
scattering. Hence, the scattering can be considered as a random process.

As the passages through the vicinity of a resonance lead to the destruction of adiabatic invariance
of $\Ffi$, the structure of streamlines becomes chaotic. Note, that the expectation of the change
of $\Ffi$ (the probability of a phenomenon times the size of a change of $\Ffi$) is of the same
order for scattering and capture. Thus both processes contribute to the stochastization equally.
There is a crucial difference between these resonant phenomena and certain other routes to chaos,
like separatrix splitting. Although the total size of the resonant zone is small with the magnitude
of the perturbation and it is localized in the space, the effect is large and global: the chaotic
domain is of the order $1$ (of the order of the total size of the system) regardless of the
magnitude of perturbation.

\section{The description of motion in the vicinity of the resonance.}

Let us derive equations of motion in the vicinity of the resonant surface, that is called a {\it
resonant zone}. Apply change of variables
\[
\left(I_1, I_2\right) \to \left(\sg, \om \right),
\]
such that
\be
\frac{\pt \sg}{\pt I_1} \frac{\pt \om}{\pt I_2} - \frac{\pt \om}{\pt I_1} \frac{\pt \sg}{\pt I_2} =
\mu(I_1, I_2),
\label{vp3}
\ee
where $\om$ is the angular velocity defined above, $\sg$ is an additional variable (chosen in such
a way that (\ref{vp3}) is satisfied) and $\mu(I_1, I_2)$ is a coefficient in the expression for the
infinitesimal volume: $d V = \mu(I_1, I_2) dI_1 dI_2 d \th$. Actually, (\ref{vp3}) can be relaxed
to be required on ${\mit R}$ only. In terms of the new variables equations of motion (\ref{vp1})
can be written as
\bea
\dot{\om} &=& \eps f_1\left( \om, \sg, \th\right),
\nonumber \\
\dot{\sg} &=& \eps f_2\left( \om, \sg, \th\right),
\label{vp4} \\
\dot{\th} &=& \om + \eps g\left( \om, \sg, \th\right). \nonumber
\eea
and the divergence-free condition has the most simple form:
\[
\frac{\pt f_1}{\pt \om} + \frac{\pt f_2}{\pt \sg} + \frac{\pt g}{\pt \th} =0.
\]
In the resonance zone we have:
\bea
\ddot{\th} &=& \eps f_1\left( \om, \sg, \th\right) + \eps \frac{\pt g}{\pt \th} \dot{\th} +
O(\eps^2)
\nonumber \\
&=& \eps \left. f_1\left( \om, \sg, \th\right)\right|_{\om=0} + \eps \left( \frac{\pt g}{\pt \th}
\dot{\th} + \frac{\pt f_1}{\pt \om}\om \right) + O(\eps^2)
\nonumber \\
&=& \eps f_1\left( 0, \sg, \th\right) + \eps \left( \frac{\pt g}{\pt \th} + \frac{\pt f_1}{\pt \om}
\right)\dot{\th} + O(\eps^2)
\nonumber \\
&=& \eps f_1\left( 0, \sg, \th \right) - \eps \frac{\pt f_2 \left( 0, \sg, \th\right)}{\pt \sg}
\dot{\th} + O(\eps^2). \nonumber
\eea
It follows from the last line in the above equation that the width of the resonant zone is given by
$\left| \dot\th \right| < \const \sqrt{\eps}$. Rescaling time,
\[
\tau = \sqrt{\eps} \; t,
\]
and denoting the derivatives with respect to $\tau$ by a prime, in the leading order (\ref{vp4})
can be written as
\bea
\sg^{\pr} &=& \sqrt{\eps} \; f_{2,0},
\label{vp5} \\
\th^{\pr\pr} &=& f_{1,0} - \sqrt{\eps} \; \frac{\pt f_{2,0}}{\pt \sg} \th^{\pr}. \nonumber
\eea
In (\ref{vp5}), we used a notation $f_{1,0} = f_1\left( 0, \sg, \th \right)$ and $f_{2,0} =
f_2\left( 0, \sg, \th \right)$. We can expand $f_{1,0}$ as
\[
f_{1,0} = a(\sg) + b(\sg, \th),
\]
where $a(\sg)$ is the average value of $f_{1,0}$ over a period:
\[
a = \frac1{2\pi} \int_0^{2\pi} f_{1,0} \: d\th,
\]
and $b(\sg, \th)$ is the remaining, zero-average part. Introduce a potential energy for $f_{1,0}$:
\[
f_{1,0} = -\frac{\pt V\left( \sg, \th\right)}{\pt \th}.
\]
The function $V$ contains a periodic and a linear in $\th$ parts:
\[
V = a(\sg) \th + \int b(\sg, \th) \; d\th = a(\sg) \th + \ovl{b}(\sg, \th).
\]
Define the resonance energy:
\be
H_R = \frac12 (\th^{\pr})^2 + V
\label{vp5a}
\ee
and denote by $h$ the value of $H_R$. Note, that at the crossing $\th^{\pr} = 0$.

The relation between $a$ and $\max\left|b(\th)\right|$ defines the properties of the phase portrait
on $(\th, \th^{\pr})$ phase plane. If
\[
\max\left|b(\th)\right| > \left| a \right|,
\]
the phase portrait looks like the one shown in Fig.~1a. For the sake of simplicity, we assumed that
there is only one oscillatory domain. The separatrix $\Sg$ is denoted by the dashed line. In the
opposite case, the phase portrait looks like the one shown in Fig.~1b.

\begin{figure}[t]
\center\epsfig{file=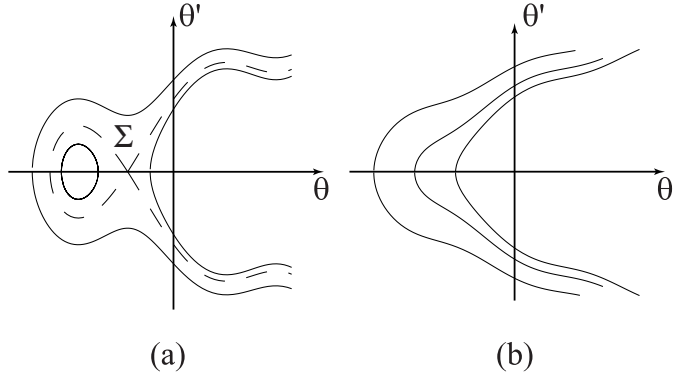, height=2in}
\label{f1}
\caption{Schematic phase portraits on the $(\th, \th^{\pr})$ plane.}
\end{figure}

\subsection{Scattering on resonance.}

The value of $\Ffi$ changes in the process of motion near and inside the resonant zone. The
cumulative change in $\Ffi$ during a single passage can be considered as a jump associated with a
resonance crossing. One can see that the theory of scattering in volume-preserving systems is the
same as in Hamiltonian systems developed in \cite{Neishtadt:1999}. We refer the reader to that
paper (see also \cite{Vainchtein:2004}) for the details of derivation and the error estimates.

Consider a streamline that crosses the resonant surface at some time moment $t_*$. Along this
streamline a phase point passes through the resonant zone fast enough, in time of order
$1/\sqrt{\eps}$. While the phase point moves inside the resonant zone changes of the slow variables
are small, of order $\sqrt{\eps}$. Therefore, for approximate description of dynamics inside the
resonant zone we can fix the value of slow variables at their values at the crossing.

Let $t_1$ and $t_2$ be two time moments, $t_1 <t_* < t_2$ such that $\left| t_{1,2} - t_*\right|
\sim 1 /\eps$ and $t_*$ is the only moment of crossing of resonance on the time interval $(t_1,
t_2)$. Then the jump of $\Ffi$ on the resonance is
\[
\Dt \Ffi = \int_{t_1}^{t_2} \dot{\Ffi} \; dt.
\]
In the first approximation, we can replace in the integral the limits $t_{2,1}$ with $\pm\infty$
(far from the resonance $\Ffi$ does not change much). We get
\be
\Dt \Ffi = \int_{-\infty}^{\infty} \dot{\Ffi} \; dt = 2 \sqrt{\eps} \int_{-s \infty}^{\th_*}
\frac{\dot{\Ffi}}{\eps} \frac1{\th^{\pr}} \: d \th = -2 s \sqrt{\eps} \int_{-s
\infty}^{\ovl{\th}_*} \dfrac{1}{\sqrt{2\left( h_* - V \right)}} \frac{\dot{\Ffi}}{\eps} \: d \th.
\label{vp5b}
\ee
where $s = \sign (a)$ and $\th_*$ and $h_*$ are the values of $\th$ and $h$, respectively, when the
exact trajectory hits ${\mit R}$. $\ovl{\th}_*$ is the resonance phase, that is equal to the
shifted value of $\th$: $\ovl{\th}_* = \ovl{\th}_* - 2\pi K$, where the value of $K$ is chosen in
such a way, that $0 < -\ovl{\th}_* - (\ovl{b}/a) \sin \ovl{\th}_* < 2 \pi$.

If the quantity $\Ffi$ can be calculated explicitly as a function of $I_1$ and $I_2$, we can
directly substitute $\dot{\Ffi} = \dot{\Ffi}(I_1, I_2, \th)$ into (\ref{vp5b}) and obtain the
expression for $\Dt \Ffi$. This is the path we take in the subsequent part of the paper, when we
consider a particular example of the resonance-induced chaotic advection in Stokes flows. However,
in many problems $\Ffi$ can only be obtained in a linearized form in the vicinity of a resonance.
In the latter case we use the change of variables (\ref{vp3}) to write (\ref{vp5b}) as
\be
\Dt \Ffi = -2 s \sqrt{\eps} \int_{-s \infty}^{\ovl{\th}_*} \dfrac{1}{\sqrt{2\left( h_* - V
\right)}} \left( \frac{\pt \Ffi(0,\sg)}{\pt \om} f_{1,0} + \frac{\pt \Ffi(0,\sg)}{\pt \sg} f_{2,0}
\right) \: d\th.
\label{vp6}
\ee
Define the quantity $\xi$ as
\[
\xi = \{ \dfrac{h_*}{2\pi \left| a \right|} \},
\]
where the curly brackets denote the fractional part. In terms of $\xi$, (\ref{vp6}) can be written
as
\be
\Dt \Ffi = -2 s \sqrt{\eps} \frac1{\sqrt{\left| a\right|}} \int_{-s \infty}^{\ovl{\th}_*}
\dfrac1{\sqrt{2\left| 2\pi\xi + \th + \ovl{b}(\th)/a \right|}} \left( \frac{\pt \Ffi(0,\sg)}{\pt
\om} f_{1,0} + \frac{\pt \Ffi(0,\sg)}{\pt \sg} f_{2,0} \right) \: d\th.
\label{vp6a}
\ee
The quantity $\xi$ characterizes a scattering. For every set of initial conditions, the values of
$\xi$ and $\Dt \Ffi$ can be calculated exactly. However, a small change of order $\eps$ in the
initial conditions produces in general a large change of order $1$ in $\xi$. Hence for small $\eps$
it is the best to treat $\xi$ as a random variable, uniformly distributed on $(0,1)$ (see
\cite{Neishtadt:1999}).

Statistical properties of the scatterings depend on the shape of the phase portrait on the $(\th,
\th^{\pr})$ plane. If the phase portrait looks like the one shown in Fig.~1a, in a generic case,
scatterings on resonance cause $\Ffi$ diffuse and drift. The rate of drift is determined by the
ensemble average of $\Dt \Ffi$ (the derivation is similar to \cite{Neishtadt:1999}, see also
\cite{Neishtadt:2005}):
\be
\left< \Dt \Ffi \right> = \int_0^1 \Dt \Ffi(\xi) d \xi = - \sqrt{\eps} \; \; \frac{\pt
\Ffi(0,\sg)}{\pt \sg} \frac1{2\pi a} \Psi(0,\sg) \sim \sqrt{\eps},
\label{vp7}
\ee
where $\Psi(0,\sg)$ is the flux of $f_{2,0}$ through the area (that we denote by $S_R$) inside the
separatrix $\Sg$ on the $(\th, \th^{\pr})$ phase plane in Fig.~1a:
\be
\Psi(0,\sg) = \int \int_{\Sg} f_{2,0} \: d\th \: d\th^{\pr}.
\label{vp8}
\ee
Note, that in Hamiltonian systems (and in many volume-preserving systems, in particular in the
system discussed below) $f_{2,0}$ does not depend on $\th$ and, therefore, $f_{2,0}$ can be taken
out of the integral and $\Psi$ becomes proportional to $S_R$.

If the phase portrait looks like the one shown in Fig.~1b, the mean change of $\Ffi$ is zero: as
there is no separatrix, $\Psi(0,\sg)=0$.

The rate of diffusion depends on the dispersion of $\Dt \Ffi$:
\[
D = \int_0^1 \left( \Dt \Ffi(\xi) - \left< \Dt \Ffi \right> \right)^2 d\:\xi \sim \eps.
\]
We return a more detailed description of the long-time dynamics in Sect.~6.

\subsection{Capture into resonance.}

The other phenomenon that affects the behavior of streamlines at a resonance crossing is capture
into resonance.

Capture is possible only if the phase portrait in the $(\th, \th^{\pr})$ plane looks like the one
shown in Fig.~1a and the phase space $(\th, \th^{\pr})$ has a well-defined separatrix.

Capture into resonance can be described as follows. The flux $\Psi(0,\sg)$ changes as a phase point
moves along a streamline. Suppose, $\Psi(0,\sg)$ increases. Then if a streamline comes very close
to the hyperbolic fixed point, it may cross the separatrix and, as a result, be caught in the
oscillatory domain within the separatrix loop. In this case, the streamline starts moving on the
resonant surface.

Captured dynamics consists of fast rotation on the $(\th, \th^{\pr})$ phase plane and slow
evolution of the parameters of the orbit, $\sg$ and $h$. The slow equations are
\bea
\sg^{\pr} &=& \sqrt{\eps} \; \left< f_{2,0} \right>,
\label{vp9} \\
h^{\pr} &=& \sqrt{\eps} \left( \left< \frac{\pt f_{2,0}}{\pt \sg} \left( \th^{\pr} \right)^2
\right> + \left< \frac{\pt V}{\pt \sg} \: f_{2,0} \right> \right), \nonumber
\eea
where $ \left< \cdot \right>$ denotes the averaging over the fast, $(\th, \th^{\pr})$, motion. The
slow dynamics is governed by a new conservation law: the flux of $f_{2,0}$ through the area inside
a closed curve, that we denote by $\Gm_R(h,\sg)$, on the $(\th, \th^{\pr})$ phase plane is
constant:
\[
\Psi (h,\sg) = \int \int_{\Gm_R(h,\sg)} f_{2,0} \; d\th \: d\th^{\pr} = \const.
\]
Recall, that the curve $\Gm_R(0,\sg)$ corresponds to the separatrix and we come back to definition
(\ref{vp8}). The captured motion is integrable and $\Psi (h,\sg)$ is the Hamiltonian.

While the streamline moves, the flux through the separatrix loop, that is a function of slow
variables, changes and, if it returns to the initial value, the streamline is released from the
resonance. If the flux keeps increasing, the once captured streamline stays captured forever (in
real settings, until it reaches the limits of the system).

As it was discussed in Sect.~3, capture can be considered as a probabilistic phenomenon: initial
conditions for streamlines that are or are not captured are mixed. Consider a point $M$ far from
the resonance such that streamline passing through $M$ intersects the resonance. Let $V^{\dt}$ be a
sphere of radius $\dt$ centered at $M$ and $V^{\dt, \eps}_c$ be the part of $V^{\dt}$ formed by
initial conditions of trajectories with a capture into the resonance (see
\cite{Neishtadt:1997,Neishtadt:1999,Itin:2000}). We define the probability of capture for the
streamlines starting at $M$ as
\[
P(M) = \sqrt{\eps} \; \lim_{\dt \to 0} \: \lim_{\eps \to 0} \frac{V^{\dt, \eps}_c/\sqrt{\eps}}{\vol
V^{\dt}}.
\]
Following \cite{Neishtadt:1999}, we have:
\be
P(M) = \sqrt{\eps} \; \frac{(\pt \Psi(0,\sg) /\pt \sg)_*}{2\pi \left|a\right|_*} \sim \sqrt{\eps},
\quad \mbox{if} \; (\pt S_R/\pt \sg)_* >0,
\label{vp9a}
\ee
where the subscript `$*$' indicates that the corresponding quantity must be evaluated at the
resonance. For $(\pt \Psi(0,\sg)/\pt \sg)_* <0$, $P(M)=0$.

\pagebreak

\section{An example: Taylor-Couette Stokes flow.}

In the rest of the paper we discuss the chaotic advection and mixing in a particular example of
volume-preserving system. Besides illustrating a general theory developed in the previous sections,
we would like to demonstrate that the resonance phenomena can be used in real settings to create
flows with coexisting mixing and non-mixing (KAM) domains with the respective size and mixing rates
easily and independently manageable. As a model base problem we chose a relatively simple system --
a Taylor-Couette flow between two infinite counter-rotating circular coaxial cylinders. Following
\cite{CartwrightFeingoldandPiro:1996}, we restrict our consideration to Stokes (small $Re$) flows.
However, we would like to stress that the whole phenomena is structurally stable, which means that
it survives most of the changes in the geometry of the flow settings and peculiarities of the flow
perturbations.

In the system under consideration, the unperturbed flow is given by (a solution to a 2-D Stokes
equation):
\bea
\dot{r} &=& 0,
\nonumber \\
\dot{z} &=& 0,
\label{vpc1} \\
\dot{\th} &=& \om\left( r,z\right), \nonumber
\eea
with
\[
\om(r,z) = \om_i \frac{r}{r_i} \left( 1- \frac{1}{1-\eta^2} \left( 1-\eta\frac{\om_o}{\om_i}
\right) \right) + \om_i \frac{r_i}{r}\left( 1- \eta\frac{\om_o}{\om_i} \right) \frac{1}{1-\eta^2},
\]
where
\[
\eta = r_i/r_o.
\]
In the above equations, $r_i$, $r_o$ are the radii of the inner and outer cylinders, respectively.
In what follows, we assume that the inner cylinder rotates with a constant angular speed $\om_i$,
while the angular speed of the outer cylinder changes periodically with $z$ around the average
value of $-\om_i$: $\om_o=\om_i(-1+\dt \sin(\la z))$, where $\la$ and $\dt$ are the wavenumber and
amplitude of oscillations, respectively. Qualitatively same results (differing in the exact
expression for the adiabatic invariant and, consequently, in specifics of the resonance equations)
can be obtained if, instead of the varying $\om_o$, the outer cylinder has a varying radius.
Similarly, we could consider a flow inside a torus-shaped container instead of the annulus.

To introduce dimensionless variables, we rescale the time by $\om_i$ and the distances by $r_i$:
\[
\ovl{t} = t \om_i, \quad \rho=\frac{r}{r_i}.
\]
From now on we assume that all the variables are in the dimensionless form and the dot denotes the
derivative with respect to $\ovl{t}$. For the numerical simulations presented below the following
set of parameters was used:
\[
r_i = 0.2, \; r_o = 1.0, \; \eta = 0.2, \; \dt = 0.4, \; \la = 2\pi.
\]
In what follows, we assume periodic boundary conditions in (normalized) $z$ with a period $1$.

The value of $\rho$ changes between $\rho=1$ (at the inner cylinder) and $\rho=1/\eta$ (at the
outer cylinder). For $\om(\rho,z)$ we have:
\be
\om(\rho,z) = -\rho \frac{\eta}{1-\eta} +\frac1{\rho} \frac{1}{1-\eta} + \frac{\eta}{1-\eta^2} \dt
\sin(\la z) \left(\rho - \frac1{\rho} \right).
\label{vpc2}
\ee
A perturbation consists of two parts. First, the inner cylinder is propagating upward (in the axial
direction) with a constant (slow) speed $\dot{z} = \eps$. Second, the shape of the outer cylinder
deviates slightly from being exactly circular. The full 3-D flow is given by
\bea
\dot{\rho} &=& \eps \kp \left( \rho - 1 \right) \cos \th,
\nonumber \\
\dot{z} &=& \eps \left( 1+ \ln \rho \: / \ln \: \eta \right),
\label{vpc3} \\
\dot{\th} &=& \om\left( \rho,z\right) - \frac1{\rho} \eps \kp \left( 2\rho- 1 \right) \sin \th.
\nonumber
\eea
In (\ref{vpc3}), $0<\eps\ \ll 1$ is a small parameter, while $\kp \sim 1$ defines a characteristic
ratio of the two perturbations. By changing the value of $\kp$ we can vary the properties of the
resonance phenomena (see Subsect.~5.3 below). One can see that the axial velocity, $\dot{z}$,
equals $\eps$ at $\rho=1$ and vanishes at $\rho=1/\eta$. A limiting case $\kp = 0$ corresponds to
the outer cylinder being circular and the axial symmetry of the flow. Note, that the reason for
choosing both perturbations to be of the same order is that it is the interaction between the
perturbations that results in the chaotic advection.

\subsection{The averaged system and the structure of the resonance.}

One can see that in (\ref{vpc3}) the variables $\rho$ and $z$ are slow and the variable $\th$ is
fast. In the first approximation we can average (\ref{vpc3}) over $\th$ to get the following
averaged equations of motion (cf. (\ref{vp2})):
\bea
\dot{\rho} &=& 0,
\nonumber \\
\label{vpc4} \\
\dot{z} &=& \eps \left( 1+ \ln \rho \: / \ln \: \eta \right). \nonumber
\eea

The averaged trajectories (in the full 3-D, $(\rho,z,\th)$, space) wind the cylinders of constant
radius with the direction of the rotation depending on the value sign of $\om$. System (\ref{vpc4})
is Hamiltonian with Hamiltonian
\be
\eps \Ffi = \eps r_i \left( \rho+ \frac1{\ln \: \eta} \left( \rho \: \ln \rho - (\rho - 1) \right)
\right).
\label{vpc4a}
\ee
The quantity $\Ffi$ is analogous to $H$ in \cite{CartwrightFeingoldandPiro:1996}. It is an integral
of the averaged system and is an adiabatic invariant of the exact system: in the absence of
resonances it would be conserved with the accuracy of order $\eps$ over times of order $1/\eps$
(see \cite{Bogolyubov:1961}).

The averaging is valid away from a resonance surface {\mit R}, where $\om = 0$, that is given by
\[
\rho_R(z) = \sqrt{\frac1{\eta} \frac{1+\eta-\eta\dt\sin(\la z)}{1+\eta-\dt\sin(\la z)}}.
\]
Trajectories that do not intersect {\mit R} remain regular. As in the averaged system the value of
$\rho$ does not change (see (\ref{vpc4})), in the first approximation, the chaotic domain occupies
the annulus between $\rho_{min}$ and $\rho_{max}$, that are the minimum and maximum values of
$\rho_R(z)$, respectively, and are given by
\[
\rho_{min} = \sqrt{\frac1{\eta} \frac{1+\eta+\eta\dt}{1+\eta+\dt}} \; \mbox{and} \; \rho_{max} =
\sqrt{\frac1{\eta} \frac{1+\eta-\eta\dt}{1+\eta-\dt}}.
\]
For the above values of the parameters, we get that the mixing domain is located between
\be
\rho_{min} = 2 \; \mbox{and} \; \rho_{max} = \sqrt{7} \apx 2.65.
\label{vpc4c}
\ee
Note, that $\rho_{min}$ and $\rho_{max}$ do not depend on the strength of perturbation, $\eps$. The
division of the flow domain on the chaotic and regular subdomains is shown in Fig.~2. The parts to
the left and to the right of the corresponding vertical lines are filled with KAM tori. One can see
that for $\dt=1$, we get $\rho_{max} = 1/\eta$ (i.e. $r_{max} = r_o$), in other words, all the
outside part of the annulus is chaotic. To mix the part adjacent to the inner cylinder, it is more
effective to modulate the frequency of the inner, not of the outer, cylinder. In the opposite
limit, as $\dt \to 0$, the chaotic domain shrinks to the zero width around $\rho = \sqrt{1/\eta}$.

\begin{figure}[t]
\center\epsfig{file=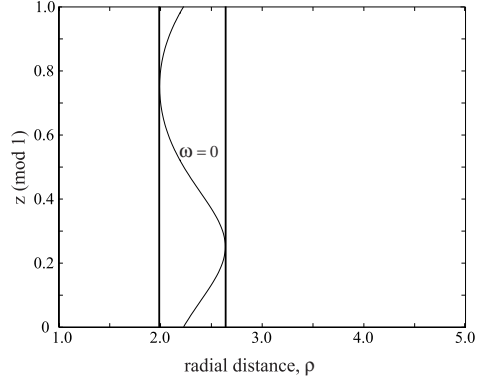, height=2in}
\label{f2}
\caption{Division of the flow domain. A chaotic domain is between the vertical lines. A regular
(KAM) domain consists of two parts at the left and at the right. The wavy line in the middle is the
resonance.}
\end{figure}

\subsection{Resonance variables.}

We can define a new variable $\sg$:
\be
\sg = - \frac{1-\eta}{2\eta} \int_0^z \frac{\rho_R(z)}{1-\frac{\dt}{1+\eta}\sin(\la z)} \; dz,
\ee
Note, that $\sg$ is just an auxiliary variable. All the final results will be given in terms of the
original variables: $\rho$ and $z$. Condition (\ref{vp3}) is satisfied on $\mit{R}$ only. As $\sg$
is a function of $z$ only, we can simplify the diffusion equations below. In the right hand side of
(\ref{vp5}) we have:
\bea
f_{1,0} &=& a + b_1 \cos \th,
\nonumber \\
\nonumber \\
a &=& \frac{\eta}{1-\eta^2} \dt \la \cos(\la z) \left(\rho - \frac1{\rho} \right) \left( 1+ \ln
\rho \: / \ln \: \eta \right),
\nonumber \\
\nonumber \\
b_1 &=& -2 \kp \frac1{\rho+1}, \nonumber
\eea
and
\[
f_{2,0} = - \frac{r_i}2 \rho \left(\rho^2-1\right) \: \left( 1+ \ln \rho \: / \ln \: \eta \right).
\]
The relation between $a$ and $b_1$ defines the properties of the phase portrait on $(\th,
\th^{\pr})$ phase plane. If
\be
\left| b_1 \right| > \left| a \right|,
\label{vpc4b}
\ee
the phase portrait looks like the one shown in Fig.~1a. If (\ref{vpc4b}) does not hold, the phase
portrait looks like the one shown in Fig.~1b. Note, that $a=0$ at $\la z = \pi/2$ and $\la z =
3\pi/2$ (where $\cos(\la z) = 0$). Therefore, while in the center of the mixing region condition
(\ref{vpc4b}) may be either satisfied or not, there are always zones near the edges of the mixing
regions, where (\ref{vpc4b}) is satisfied.

\subsection{Scattering.}

In the problem under consideration, we have an explicit form of $\Ffi = \Ffi(\rho)$. Hence, we can
use equation (\ref{vp5b}) to calculate the jumps $\Dt \Ffi$ (or $\Dt \rho$). We have:
\be
\Dt \Ffi = \int_{-\infty}^{\infty} \dot{\Ffi} \: d t = -2 s \sqrt{\eps} \: r_i \kp \left( \rho-1
\right) \left( 1+ \ln \rho \: / \ln \: \eta \right) \int_{-s \infty}^{\ovl{\th}_*} \dfrac{\cos
\th}{\sqrt{2\left( h_* - V \right)}} \: d\th,
\label{vpc5}
\ee
where $s = \sign \: (a)$. The value of $\rho$ must be taken at the moment of crossing, $\rho =
\rho_*$, and the resonance potential, $V$, is
\[
V(\sg, \th) = -a \th - b_1 \sin \th.
\]
In terms of $\xi$, we can write (\ref{vpc5}) as
\[
\Dt \Ffi = -2 s \sqrt{\eps} \kp \: r_i \frac{\left(\rho-1 \right)}{\sqrt{\left| a\right|}} \left(
1+ \ln \rho \: / \ln \: \eta \right) \int_{-s \infty}^{\ovl{\th}_*} \dfrac{\cos \th}{\sqrt{2\left|
2\pi\xi + \th + (b_1/a) \sin \th \right|}} \: d\th,
\]
with
\[
\xi = \{ \dfrac{h_*}{2\pi \left| a \right|} \} = \frac1{2\pi} \left| - \ovl{\th}_* -  (b_1/a) \sin
\ovl{\th}_* \right|.
\]
The quantity $\ovl{\th}_*$ was defined in Sect.~4.1. Statistical properties of the scatterings
depend on the shape of the phase portrait on the $(\th, \th^{\pr})$ plane. If the phase portrait
looks like the one shown in Fig.~1a, the ensemble average of $\Dt \Ffi$ is (see Sect.~4):
\[
\left< \Dt \Ffi \right> = - 2 s \sqrt{\eps} \; \kp r_i \frac{\left( \rho-1 \right)}{2\pi b_1}
\left( 1+ \ln \rho \: / \ln \: \eta \right) S_R = - s \sqrt{\eps} \; \frac{\rho^2 - 1}{2\pi} \left(
1+ \ln \rho \: / \ln \: \eta \right) S_R,
\]
where $S_R$ is the area under the separatrix loop in Fig.~1a:
\[
S_R = 2 \int_{\th_{min}}^{\th_{max}} \sqrt{-2 (V - V_c)} d \th,
\]
and $V_c$ is the value of $V$ at the hyperbolic fixed point in Fig.~1a:
\[
V_c = -a \th_c - b_1 \sin \th_c; \quad \cos \th_c = -a/b_1.
\]
If the phase portrait looks like the one shown in Fig.~1b, $\left< \Dt \Ffi \right> = 0$: as there
is no separatrix, $S_R=0$.

As $\Ffi$ is a function of $\rho$ only (see (\ref{vpc4a})), we can use $\rho$ instead of $\Ffi$. We
have:
\be
\Dt \rho = -2 s \sqrt{\eps} \: \kp \frac{ \rho - 1}{\sqrt{\left| a\right|}} \int_{-s
\infty}^{\ovl{\th}_*} \dfrac{\cos \th}{\sqrt{2\left| 2\pi\xi + \th + (b_1/a) \sin \th \right|}} \:
d\th.
\label{vpc7a}
\ee
Equation (\ref{vpc7a}) was checked numerically for various values of parameters $\xi, \kp$ and
$\eps$. In Fig.~3 the plots of $\Dt \rho \left( \xi \right)/\sqrt{\eps}$ are presented for (a) $\kp
= 2$ (when (\ref{vpc4b}) is satisfied) and (b) $\kp=0.2$, (when (\ref{vpc4b}) is not satisfied).
The solid lines in Fig.~3 correspond to theoretical values of $\Dt \rho \left( \eps
\right)/\sqrt{\eps}$ given by (\ref{vpc7a}), and the asterisks show values obtained numerically
from (\ref{vpc1}) for various values of $\xi$. When inequality (\ref{vpc4b}) is satisfied, $\Dt
\rho(\xi)$ has a singularity. Therefore, there is a possibility (albeit quite small) of large
changes in $\rho$ in the process of scattering. However, as this singularity is logarithmic,
$\left< \Dt \rho \right>$ is finite:
\be
\left< \Dt \rho \right> = - s \sqrt{\eps} \; \frac{\rho^2 - 1}{2\pi} S_R .
\label{vpc7}
\ee
The plot $\left| \left< \Dt \rho \right> \right|/\sqrt{\eps}$ as a function of $\rho$ for $\kp=2$
(the values of other parameters defined in (\ref{vpc4c})) is shown in Fig.~4. Note, that the value
of $S_R$ (and, therefore, of $\left< \Dt \rho \right>$) depends crucially on the $\kp$ and $\dt$.
In particular, $S_R \to 0$ as $\kp \to 0$.

Note, that the values of $\rho$ in the right hand side of (\ref{vpc7a}) and (\ref{vpc7}) are,
strictly speaking, those at the moment of crossing. But, as characteristic values of $\Dt \rho$ at
a single crossing are small (it is the accumulation of those changes that leads to the mixing), if
we specify the initial conditions far from $R$, we can use those values of $\rho$ in (\ref{vpc7a})
and (\ref{vpc7}).

It follows from the definitions of $s$ and $a$, that $\left< \Dt \rho \right> >0$ if $\cos(\la
z)<0$ and $\left< \Dt \rho \right> <0$ if $\cos(\la z)>0$. As every single jump is small, two
consecutive crossings occur at almost the same values of $\rho$ and $S_R$. Therefore, it follows
from (\ref{vpc7}), that the average change in $\rho$ during one $z$-period is zero.

\begin{figure}[t]
\center\epsfig{file=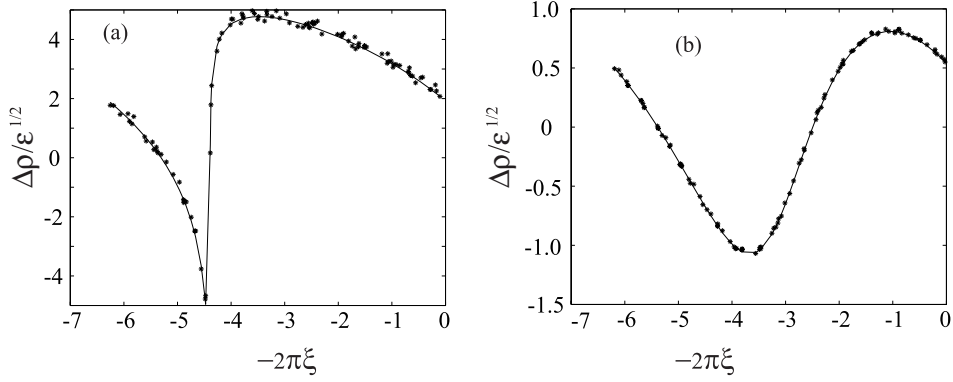, height=2in}
\label{f3}
\caption{The plot of $\Dt \rho/\sqrt{\eps}$ as a function of $\xi$; (a): $\kp = 2$ and (b):
$\kp=0.2$. Note the difference in scales.}
\end{figure}

\begin{figure}[t]
\center\epsfig{file=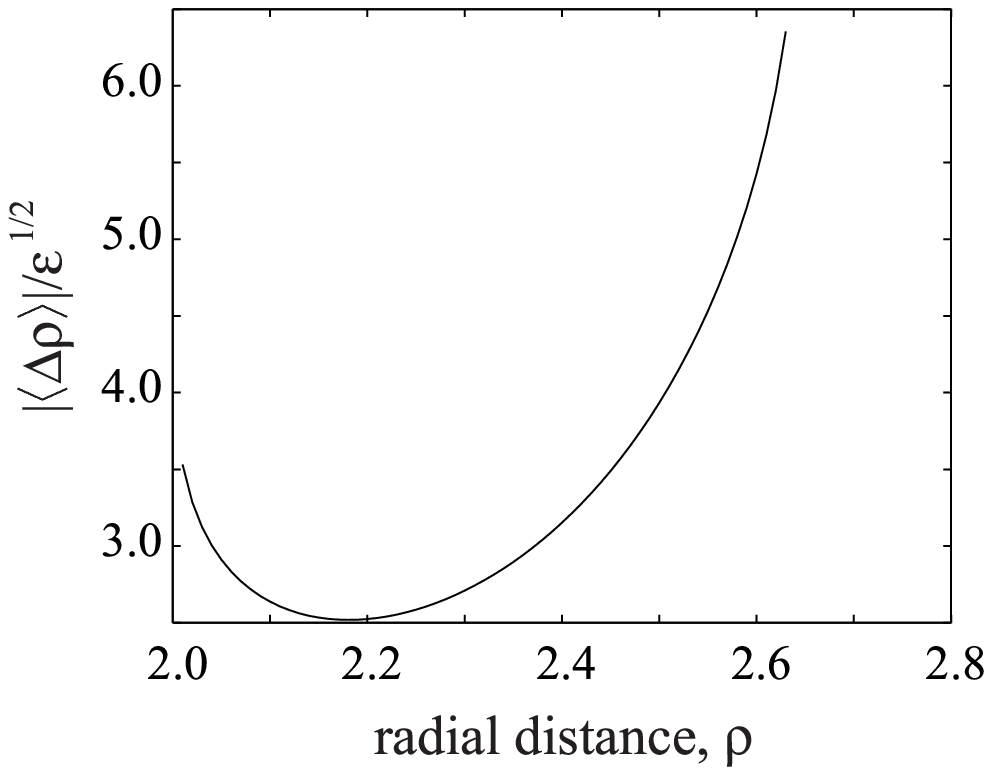, height=2in}
\label{f4}
\caption{The plot of $\left| \left< \Dt \rho \right> \right|/\sqrt{\eps}$ as a function of $\rho$
for $\kp=2$ and the values of other parameters defined in (\ref{vpc4c}).}
\end{figure}

\subsection{Capture.}

A small percentage of particles do not leave the vicinity of resonance quickly, but are captured
into resonance and travel near $R$ for a long, of order of $\sim 1/\eps$, time.

It was discussed in Sect.~4, that capture can be considered as a probabilistic process with a
probability to be captured from a small ball of initial conditions given by (\ref{vp9a}):
\be
P(M) = \sqrt{\eps} \; \frac{(\pt S_R/\pt \sg)_*}{2\pi \left|a\right|_*} = \sqrt{\eps} \;
\frac1{2\pi \left|a\right|_*} \left(\frac{\pt S_R}{\pt z}\right)_* \left(\frac{\pt z}{\pt
\sg}\right)_* \sim \sqrt{\eps}
\label{vpc9a}
\ee

A typical captured dynamics (for $\eps = 10^{-4}$ and $\kp = 2$) is shown in Fig.~5. A projection
on the slow, $(\rho,z)$, plane and the time evolution of $\om(\rho,z)$ are shown in Figs.~5a and
5b, respectively. A streamline comes from the bottom in Fig.~5a (from the left in Fig.~5b), is
captured near $z=0.05$ ($t=100$), moves along the resonance, is released from the resonance near
$z=0.45$ ($t=1000$), and then proceeds along an adiabatic path. To describe the captured dynamics
we could use equations (\ref{vp9}) with expressions for $f_{2,0}$ and $V$ derived in the previous
subsection. But as $f_{2,0}$ does not depend on $\th$, the evolution equation for $\sg$,
\[
\sg^{\pr} = \sqrt{\eps} \; f_{2,0}(\sg),
\]
can be solved separately. As $\sg$ depends on $z$ only, we can return to the original variables to
get
\be
z^{\pr} = \sqrt{\eps} \left( 1+ \ln \rho \: / \ln \: \eta \right)_{\om =0} = \sqrt{\eps} \left( 1+
\frac1{2\ln \: \eta} \; \ln \left( \frac1{\eta} \frac{1+\eta-\eta\dt\sin(\la z)}{1+\eta-\dt\sin(\la
z)} \right) \right).
\label{vpc9}
\ee
Note, that equation for the captured motion (\ref{vpc9}) can be solved once regardless of the
initial conditions. Different initial conditions (i.e. different moments of capture) can be taken
into account by shifting the origin of the slow (captured) time. The averaged (over fast $(\th,
\th^{\pr})$ oscillations) evolution $\rho(\tau)$ is given by the resonance condition
$\om(\rho,z(\tau)) = 0$. It follows from the symmetry $\la z \to \pi-\la z$ that the release occurs
at approximately the same value of $\rho$ (with the accuracy of order of $\sqrt{\eps}$) at which
the capture happened. It takes a phase point time (recall, that the original time differs from the
slow time $\tau$ by a factor $1/\sqrt{\eps}$)
\[
T_c = \frac1{ \sqrt{\eps}} \: \int_{z_{cap}}^{z_{rel}} \; \frac1{z^{\pr}} \: d z
\]
to travel from capture to release.

\begin{figure}[t]
\center\epsfig{file=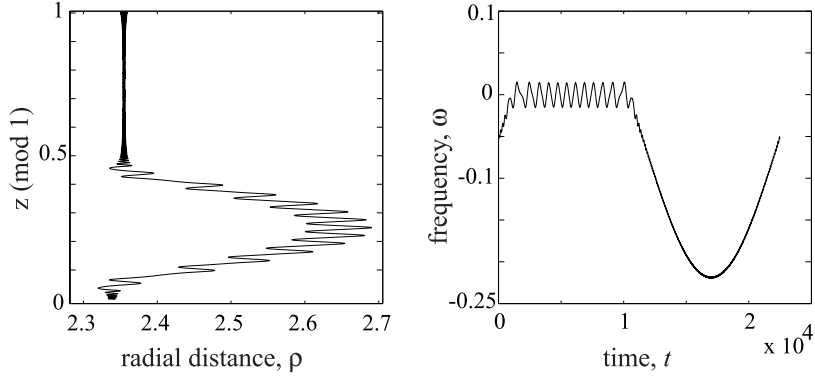, height=2in}
\label{f5}
\caption{Captured motion. (a) A projection of a streamline on the slow, $(\rho,z)$, plane; and (b)
the time evolution of $\om(\rho,z)$.}
\end{figure}

\section{Long term dynamics and numerical simulations.}

It was shown before (see e.g. \cite{Vainchtein:2004} and references therein) that in the
Hamiltonian systems the accumulation of jumps of an adiabatic invariant at resonances leads to the
chaotic advection and mixing. In the present section we study the dynamics over the long intervals
of time that include many (of order of $\sim 1/\eps$) crossings.

There are two quantities that describe the chaotic advection: the size of the chaotic domain and a
characteristic rate of mixing inside the chaotic domain.

It was shown in Sect.~5, that the chaotic domain is a cylinder between $r_{min}$ and $r_{max}$.
Outside the chaotic domain the majority of streamlines are regular. More precisely, they are
regular if they are removed from the resonance by a distance that is small with $\eps$. A technique
to estimate the excess width of the resonance domain was developed in \cite{VVN:1996b}. A
projection of three representative streamlines on the slow, $(r,z)$, plane is shown in Fig.~6.
Almost straight vertical lines are regular streamlines. A single streamline that starts at
$r_{in}=0.45$ fills almost the entire chaotic domain.

\begin{figure}[t]
\center\epsfig{file=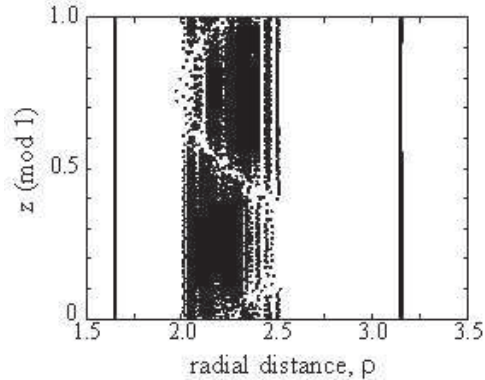, height=2in}
\label{f6}
\caption{Regular and chaotic domains. Almost straight vertical lines are regular streamlines. A
single streamline that starts at $r_{in}=0.45$ fills almost the entire chaotic domain.}
\end{figure}

Note, that the size and the location of the chaotic domain depends on the size of the annulus,
$\eta$, and the amplitude of the frequency oscillations, $\dt$, and is independent of the magnitude
of perturbation, $\eps$. The size of the chaotic domain is of order $1$ (i.e. it is on the scale of
the whole system) regardless of how small $\eps$ is. This property sets the resonance phenomena
(and separatrix crossings, see e.g. \cite{VVN:1996a}) aside from most of the other "routes to
chaos." While the phenomena itself is local (we need to solve equations of motion only in the
vicinity of a resonance surface), the effect is global and does not disappear as $\eps$ becomes
smaller and smaller.

Recall that for $\eps=0$ the dynamics is regular (see Sect.~5.1) and we may expect a smooth
transition from partially chaotic to fully regular dynamics as $\eps \to 0$. The only chance to
achieve it is if as $\eps \to 0$ it takes longer and longer time for streamlines to mix. In other
words, the rate of mixing (diffusion) must go to zero, as $\eps \to 0$. To estimate the rate of
diffusion we need to know the statistics of the changes of the adiabatic invariant. In a generic
case, depending on the ratio between the coefficients in $V$ (the resonance potential), the value
of $\Ffi$ may drift or diffuse. However, in the particular case of the flow given by (\ref{vpc1}),
it follows from the symmetry $\la z \to \pi-\la z$, that $\left< \Dt \rho \right>$ at two
consecutive crossings cancel each other. Therefore, the statistics of the changes of $\rho$ is
diffusive regardless of the validity of (\ref{vpc4b}).

Quantitative properties of the diffusion of adiabatic invariant depend on whether consecutive
crossings are statistically dependent or independent. Statistical independence follows from the
divergence of phases $\th$ along trajectories and for Hamiltonian systems was discussed in
\cite{Neishtadt:1999}. In volume-preserving systems it can be illustrated following the similar
lines (see also \cite{Dolgopyat:2004}). Assuming the statistical independence of consecutive
crossings, we can estimate the rate of mixing. The evolution of $\rho$ can be considered as a
random walk with a characteristic step of order of $\sim \sqrt{\eps}$. Hence, after $N$ crossings a
value of $\rho$ changes by a quantity of order of $\sim \sqrt{N} \times \sqrt{\eps}$. The mixing
can be called complete if a trajectory spreads all over a chaotic domain. The difference between
the values of $\rho$ that bound the chaotic domain in our problem is of order $1$. Therefore, it
takes approximately $N \sim 1/\eps$ resonance crossings to complete the diffusion. As a typical
time between successive crossings is of order of $\sim 1/\eps$, a characteristic time of mixing,
$T_M$, is of order of $\sim \eps^{-2}$. One can see that, like we suggested above, a characteristic
time of mixing goes to infinity as $\eps \to 0$ and the rate of mixing, defined as $1/T_M$,
vanishes as $\eps \to 0$. The similar estimate was reported in
\cite{FeingoldandKadanoffandPiro:1988}.

Another important limiting case corresponds to $\kp \to 0$. For $\kp = 0$ the outer cylinder is
circular and we regain the axial symmetry. In this case the resonances do not lead to chaotic
advection. Indeed, it follows from (\ref{vpc7a}), that a characteristic size of the jumps $\Dt \rho
\to 0$ as $\kp \to 0$.

To check to the validity of theory developed in the previous sections, we performed a set of
numerical simulations. We used the set of parameters specified in the beginning of Sect.~6 and
\[
\eps=10^{-6}; \; \kp=2.
\]

First, we checked the validity of formula for a jump in adiabatic invariant at a single crossing
(Eq. (\ref{vpc7a})). Those results were presented in Sect.~6.3. Second, we studied the diffusion of
adiabatic invariant and large scale mixing. For that purpose we took $1000$ initial conditions that
are uniformly distributed in a box of $\rho_{in} \times z_{in} \times \th_{in} = [2.249, 2.251]
\times [-0.01, 0.01] \times [-0.01, 0.01]$ and considered the Poincare sections located at
$z=N+0.25$ and $z=N+ 0.75$, where $N$ is a set of integer numbers. Every trajectory crosses the
resonance once between the consecutive sections. However, to simplify the following, it is
convenient to consider a change in $\rho$ after two successive crossings, $\Dt \rho_{\Sg} = \Dt
\rho_{0.25 \to 0.75} + \Dt \rho_{0.75 \to 0.25}$, as $\Dt \rho_{\Sg}$ has zero mean.

Let $\Psi(\rho,N)$ denote the number of trajectories that, after $N$ double crossings, have the
value of $\rho$ between $\rho-0.005$ and $\rho+0.005$. The spreading of $\Psi(\rho,N)$, obtained by
integrating (\ref{vpc1}) with the initial conditions specified above, is shown in Fig.~7.
\begin{figure}[t]
\center\epsfig{file=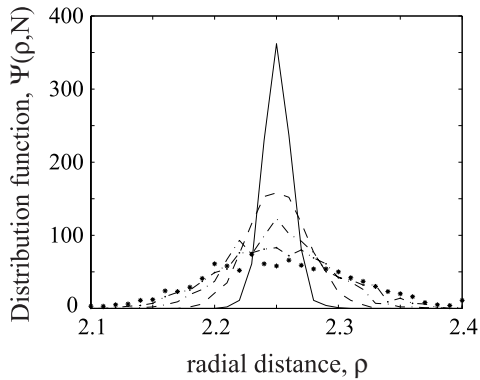, height=2in}
\label{f7}
\caption{The histogram of $\Psi(\rho,N)$ after different numbers of the double resonance crossings.
The solid line and the stars correspond to $N=10$ and $N=400$, respectively. The other curves are
between those values of $N$.}
\end{figure}

Many systems with random walks are described using diffusion equations, that for a system under
consideration can be written as
\be
\frac{\pt \Psi_D}{\pt t} = \frac{\pt }{\pt \rho} \left( D(\rho) \: \frac{\pt \Psi_D}{\pt \rho}
\right),
\label{vpc10}
\ee
where $\Psi_D$ is a probability distribution function. To relate Eq.~(\ref{vpc10}) to the evolution
of $\Psi(\rho,N)$, the time $t$ should be measured in the units of the period in the $z$-direction
(in other words, there is one double resonance crossing per unit time). The diffusion coefficient
$D(\rho)$, that we call the coefficient of the {\it adiabatic diffusion}, is given by the
dispersion of $\Dt \rho_{\Sg}$:
\[
D = D(\rho; \eps, \kp, \la) = \int_0^1 \left( \Dt \rho(\xi) - \left< \Dt \rho \right> \right)^2
d\:\xi .
\]
In the leading order we have
\[
D(\rho; \eps, \kp, \la) = \eps \wtd{D}(\rho; \kp, \la).
\]
The profile of the normalized dispersion, $\wtd{D}$, as a function of $\rho$ for the values of the
parameters specified above is presented in Fig.~8.
\begin{figure}[t]
\center\epsfig{file=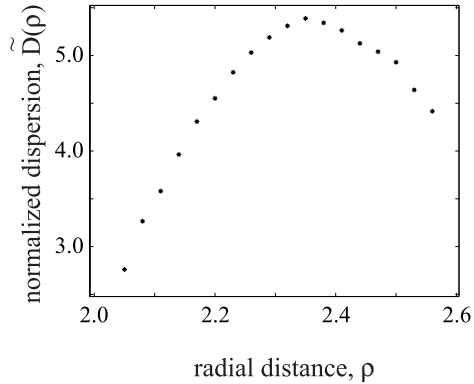, height=2in}
\label{f8}
\caption{The normalized dispersion, $\wtd{D}$, as a function of $\rho$.}
\end{figure}
The second moment of $\Psi(\rho,N)$, is presented in Fig.~9. The constant slope confirms the
diffusion assumption and the magnitude of the slope is in the good agreement with the prediction
based on the simplified diffusion equation with $D(\rho) = D(\rho=2.25) \apx 5 \times 10^{-6}$.
\begin{figure}[t]
\center\epsfig{file=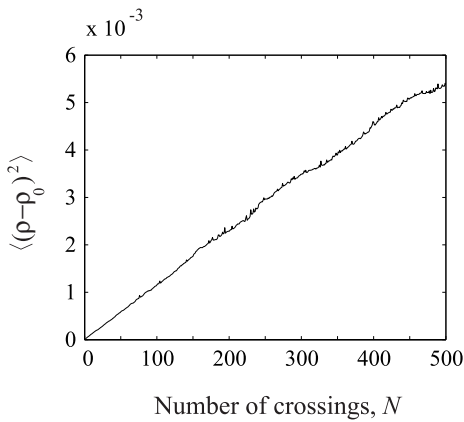, height=2.2in}
\label{f9}
\caption{The square of the standard deviation, $\left< \left(\rho - \rho_0\right)^2 \right>$,
averaged over $1000$ trajectories as a function of the number of the double resonance crossings,
$N$.}
\end{figure}

The above considerations describe the diffusion across the level of constant adiabatic invariant
(in radial direction). Mixing in the axial direction is due to the radial gradient in the axial
velocity, while mixing in the azimuthal direction is due to the gradient in $\om(\rho,z)$. Both
processes are much faster then the radial diffusion and by the time of order of $1/\eps^2$ the
distribution in those directions is quite uniform.

\section*{Conclusions and Acknowledgements}

In the present paper we considered capture into resonance and scattering on resonance in 3-D
volume-preserving slow-fast systems. In the first sections we proposed a general theory of those
processes. Then we applied it to a class of viscous Taylor-Couette flows between two
counter-rotating cylinders. We described the phenomena during a single passage through resonance
and showed that multiple passages resulted in the chaotic advection and mixing. We calculated the
width of the mixing domain and estimated a characteristic time of mixing. We showed that in the
systems with resonances  mixing can be described using a diffusion equation with a diffusion
coefficient depending on the averaged effect of the passages through resonances. Finally, we would
like to stress that the resulting diffusion coefficient is not related to the molecular diffusion.

D.V. and A.N. are grateful to Russian Basic Research Foundation Grant No. 03-01-00158. D.V. and
I.M. are grateful to AFOSR Grant No. F49620-03-1-0096 and ITR-NSF Grant ACI-0086061.

\bibliography{refer1}

\end{document}